\documentclass[pdflatex,sn-mathphys-num]{sn-jnl}% Math and Physical Sciences Numbered Reference Style
\usepackage{graphicx}%
\usepackage{multirow}%
\usepackage{amsmath,amssymb,amsfonts}%
\usepackage{amsthm}%
\usepackage{mathrsfs}%
\usepackage[title]{appendix}%
\usepackage{xcolor}%
\usepackage{textcomp}%
\usepackage{manyfoot}%
\usepackage{booktabs}%
\usepackage{algpseudocode}%
\usepackage{listings}%
\usepackage{color}
\usepackage{tabularray}
\usepackage{caption}        
\usepackage{tcolorbox}
\usepackage{xurl}            % allow URL line breaks (best effort)
\usepackage{hyperref}
\hypersetup{breaklinks=true} % allow link breaks
\usepackage[ruled,vlined]{algorithm2e}
\tcbset{
  colback=white,
  colframe=black,
  boxrule=0.6pt,      
  arc=0pt,           
  left=4pt,right=4pt, 
  top=3pt,bottom=3pt,
  fontupper=\footnotesize\itshape
}
\usepackage[dvipsnames]{xcolor}
\tcbset{
  rqstyle/.style={
    colback=gray!12,     % 灰底深浅可调：gray!10 ~ gray!18
    colframe=black,      % 黑色边框
    boxrule=0.9pt,       % 边框粗细
    arc=3pt,             % 圆角程度（截图风格）
    left=6pt,right=6pt,
    top=4pt,bottom=4pt,
    boxsep=0pt,
    before skip=6pt,
    after skip=6pt
  }
}

\newenvironment{rqanswer}[1]{%
  \begin{tcolorbox}[rqstyle]
  \small\noindent\textbf{Answer to #1: }%
}{%
  \end{tcolorbox}
}

\theoremstyle{thmstyleone}%
\theoremstyle{thmstyletwo}%

\theoremstyle{thmstylethree}%

\begin{document}

\title[Article Title]{Are LLMs Reliable Code Reviewers? Systematic Overcorrection in Requirement Conformance Judgement}

\author{Haolin Jin}\email{hjin3177@uni.sydney.edu.au}
\author{Huaming Chen}\email{huaming.chen@sydney.edu.au}

\affil{\orgdiv{University of Sydney}, \orgaddress{\city{Sydney}, \state{NSW}, \country{Australia}}}

%%==================================%%
%% Sample for unstructured abstract %%
%%==================================%%

\abstract{Large language models (LLMs) have become essential tools in software development, widely used for requirements engineering, code generation and review tasks. Software engineers often rely on LLMs to verify if code implementation satisfy task requirements, thereby ensuring code robustness and accuracy. However, it remains unclear whether LLMs can reliably determine code against the given task descriptions, which is usually in a form of natural language specifications. 
In this paper, we uncover a systematic failure of LLMs in matching code to natural language requirements. Specifically, with widely adopted benchmarks and unified prompts design, we demonstrate that LLMs frequently misclassify correct code implementation as non-compliant or defective. Surprisingly, we find that more detailed prompt design, particularly with those requiring explanations and proposed corrections, leads to higher misjudgment rates, highlighting critical reliability issues for LLM-based code assistants.
We further analyze the mechanisms driving these failures and evaluate the reliability of rationale-required judgments. Building on these findings, we propose a Fix-guided Verification Filter that treats the model proposed fix as executable counterfactual evidence, and validates the original and revised implementations using benchmark tests and spec-constrained augmented tests. Our results expose previously under-explored limitations in LLM-based code review capabilities, and provide practical guidance for integrating LLM-based reviewers with safeguards in automated review and development pipelines.}


%%\pacs[JEL Classification]{D8, H51}

%%\pacs[MSC Classification]{35A01, 65L10, 65L12, 65L20, 65L70}

\maketitle
\section{Introduction}
As large language models (LLMs) have demonstrated increasing capability in the domain of code synthesis \cite{austin2021program}, a growing body of research and tooling efforts have begun exploring their application for automated code review and verification \cite{xu2022systematic,rasheed2024ai,cai2025automated}. Traditional code reviews require developers to manually verify the alignment between code logic and requirements, a process that is both time consuming and prone to human error. LLMs show significant potential to reduce this burden by automating code assessments and suggesting improvements during the code-review process. For instance, recent studies have leveraged models such as GPT-4o to evaluate code submissions and determine whether they meet quality standards or require revisions \cite{liu2023your}.

In software engineering, verifying and validating that source code aligns with its task requirements remains a challenge \cite{shankar2024validates}. Requirements engineering is widely recognized as crucial to clearly defining, understanding, and aligning project objectives with stakeholder needs, thereby reducing risks associated with errors, delays, and project failures \cite{couder2024requirements}. Recent studies start to investigate the potential of LLMs to bridge this gap. Advanced models such as GPT-4o have shown promising accuracy in identifying unmet requirements from textual descriptions \cite{reinpold2024exploring}, indicating their potential as reliable ``virtual reviewers'' even without test cases or formal implementations. To further improve review effectiveness, researchers have explored self-critical mechanisms within LLMs, such as the Self-Refine framework, where models iteratively critique and refine their outputs without additional training \cite{madaan2023selfrefineiterativerefinementselffeedback}. Ideally, LLMs should accurately understand functional requirements described in natural language and reliably judge whether provided code satisfies these requirements, assisting developers in identifying defects or confirming correctness. However, in scenarios lacking test cases or reference implementations, the reliability of LLMs in performing such ``description-to-code'' evaluations remains unclear. 

A common intuition is that richer prompting improves reliability: asking the LLM to explain its decision, enumerate mismatches, or propose fixes should encourage more careful reasoning \cite{white2023prompt}. Nevertheless, LLMs are known to exhibit hallucinations and systematic biases \cite{ji2023survey}, and prompt ``enhancements'' may unintentionally shift the model's error profile rather than uniformly improving accuracy. In particular, more elaborate prompts can amplify an LLM's tendency to over-criticize correct code or to rationalize incorrect decisions with persuasive-but-unfaithful explanations. This motivates a key question for automated review: \emph{Do prompt variants trade false rejection for false acceptance, and are the generated rationales themselves trustworthy?}

To answer these questions, we investigate whether LLMs can correctly determine code correctness when provided with precise task descriptions and correct implementations. Specifically, we conduct a large-scale empirical investigation across three widely used programming benchmarks under three prompting approaches. We evaluate five representative LLMs (three closed-source and two open-source) and report full confusion-matrix statistics. Beyond decision accuracy, we further examine the reliability of the explanation along two complementary dimensions. Finally, we proposed a filter-embedded framework to assess their usefulness in reducing such bias. Our experiments reveal a concerning phenomenon: LLMs frequently issue false negative judgments, incorrectly concluding that correct implementations fail to meet stated requirements, resulting in a high false negative rate. Surprisingly, our extended experiments involving various prompt designs indicate that increasing prompt complexity, such as requiring explicit explanations and suggested corrections, counterintuitively leads to higher rates of misjudgment, this finding contradicts the common assumption that incorporating explanatory steps typically enhances the reasoning and accuracy of LLMs \cite{yu2023towards,chu2023survey}. Instead, detailed prompts may inadvertently introduce biases toward excessive fault finding, causing models to detect non-existent errors in otherwise correct implementations.
\begin{figure}
    \centering
    \includegraphics[width=\linewidth]{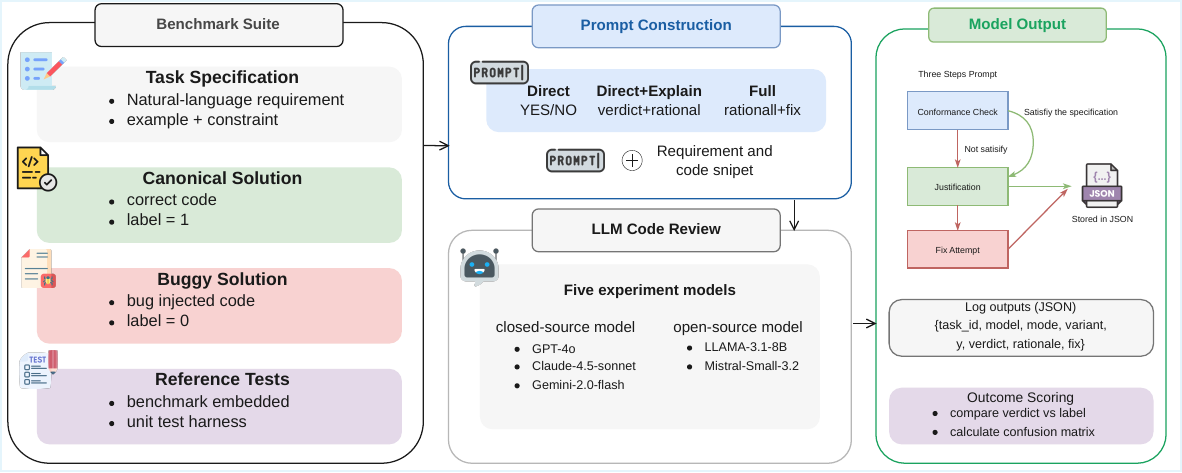}
    \caption{Workflow of evaluating LLM code conformance on canonical and buggy solutions under three prompting modes, with outputs logged for downstream scoring.}
    \label{fig:method}
\end{figure}

Our findings have direct implications for deploying LLMs as automated reviewers: even when an LLM can frequently ``pass'' correct solutions, it may simultaneously suffer from severe over-correction (high FN) and unsafe acceptance (high FP), and the accompanying explanations may not faithfully justify the decision. These limitations can undermine the utility and trustworthiness of LLM-based review systems \cite{collante2025impact,almeida2024aicodereview}, especially in automated pipelines where a model's judgement may trigger or block downstream actions \cite{panickssery2024llm}, unreliable assessments will compromise the overall effectiveness of automated engineering workflow. Therefore, understanding why LLMs systematically misjudge correct implementations and identifying strategies to mitigate these failures becomes crucial, by providing these novel insights, we aim to raise community awareness regarding the limitations of LLMs in code evaluation. In summary, this work makes the following contributions:
\begin{enumerate}
    \item \textbf{False negatives discovery} - We reveal that LLMs frequently misjudge correct code as failing to meet requirements, indicating their bias towards over-correction rather than accurate verification.
    \item \textbf{Bias characterization across prompts and five models} - We quantify over-correction bias and false acceptance under three prompting approaches and five LLMs, revealing prompt-induced tradeoffs and decision instabilities.
    \item \textbf{Self-consistency and Fault-awareness} - We introduce and evaluate an explanation-centric consistency measure that captures contradiction between model verdicts and their rationales, also we evaluate whether LLM rationales reflect fault-aware reasoning, and analyze how false acceptance concentrates on specific bug types.
    \item \textbf{Exploratory mitigation analysis} - We explore mitigation strategies to assess their effectiveness in reducing judgement biases.
\end{enumerate}

\section{Background and Related Work}
\subsection{LLMs for code review}
Modern code review is a cornerstone of current software development, serving as a primary quality check for defect discovery, maintainability, and knowledge transfer. Empirical work has characterized both the expected benefits and the practical frictions of review, highlighting that “review quality” depends not only on correctness detection but also on how reviewers interpret requirements and communicate actionable feedback \cite{bacchelli2013expectations,rigby2013convergent,mcintosh2014impact}. Against this backdrop, learning-based code intelligence has rapidly evolved from task specific neural models to general purpose pre-trained models for code and natural language. Encoder based and encoder-decoder models such as CodeBERT \cite{feng2020codebert}, GraphCodeBERT \cite{guo2021graphcodebert}, and CodeT5 \cite{wang2021codet5} established strong baselines for code understanding and generation tasks (e.g., summarization, translation, and defect-related reasoning), and provide the representational foundations that later large language models build upon. These capabilities enable an emerging class of LLM-assisted reviewing scenarios, where models are used to summarize changes, propose edits, and explain behavioral implications as part of review.

More recently, code specialized LLMs have demonstrated strong synthesis and editing capabilities, motivating their use as “virtual reviewers” that can generate explanations and suggest repairs. Representative systems and models include Codex style evaluation on HumanEval \cite{chen2021evaluating}, competition-level synthesis with AlphaCode \cite{li2022competition}, and open code LLMs such as CodeGen, InCoder, and CodeLlama \cite{nijkamp2022codegen,fried2022incoder,roziere2023code}. In parallel, software engineering benchmarks targeting real-world issue resolution (e.g., SWE-bench) further illustrate the community’s interest in using LLMs for end-to-end debugging and patching workflows \cite{jimenez2024swebench}. Our work connects to this trajectory but targets a distinct point in the review pipeline: requirement-conformance judgement without executing tests, and the systematic biases that arise when prompts elicit explanations and fixes.

\subsection{Reliability issues in code tasks}
A substantial body of work has evaluated LLM capability in code tasks via curated benchmarks with executable tests or reference outputs, such as HumanEval, MBPP, APPS, MultiPL-E, and QuixBugs \cite{chen2021evaluating,austin2021programsynthesislargelanguage,hendrycks2021measuring,cassano2022multipl,ye2021comprehensive}. These benchmarks have been instrumental in quantifying functional correctness under test-based evaluation and enabling fair comparisons across models. However, they also highlight a practical gap: many real code review settings do not provide comprehensive tests or formal specifications, and “correctness” must be assessed from a natural-language requirement and code alone. In such settings, the key risk is not only incorrect code generation, but also unreliable code assessment, models may reject correct implementations (false rejection) or accept buggy ones (false acceptance) when reasoning is unconstrained by execution.

Reliability concerns in LLM-based reasoning systems have been widely discussed in terms of hallucination, overconfidence, and specification drift, where models produce plausible but unsupported claims \cite{ji2023survey,liang2022helm,openai2023gpt4}. For code-related tasks, these risks are amplified because models can (i) infer non-existent constraints, (ii) speculate about runtime failures without evidence, or (iii) overfit to stylistic “best practices” that are orthogonal to requirement satisfaction. Moreover, the security community has shown that LLM-generated code can introduce vulnerabilities, which raises the stakes of inaccurate review judgements in practice \cite{pearce2022asleep}. 
% Together, these findings motivate measuring two-sided reliability and understanding when explanations genuinely reflect grounded reasoning versus post-hoc rationalization.

Prompting and self-improvement strategies have been proposed to increase reasoning quality, including chain-of-thought prompting, zero-shot “think step-by-step” prompting, and self-consistency sampling \cite{wei2022chain,kojima2022zeroshot,wang2022selfconsistency}. Iterative refinement and agentic approaches (e.g., Self-Refine and Reflexion) further attempt to improve outputs through critique and feedback cycles \cite{madaan2023self,shinn2023reflexion}. Related prompting frameworks such as ReAct and prompt-pattern catalogs systematize how to elicit reasoning and tool use \cite{yao2023reactsynergizingreasoningacting,white2023prompt}. While these approaches can improve performance in many settings, they also introduce new failure modes (e.g., longer rationales that drift from the requirement, or “fixes” that reflect overly conservative assumptions).

\subsection{LLMs judge and bias}
LLM-based evaluators (“LLM-as-a-judge”) have become a common mechanism for scalable assessment of model outputs, particularly in open-ended generation where reference-based metrics are weak \cite{gu2024survey}. Benchmarking frameworks such as MT-Bench and Chatbot Arena operationalize pairwise preference evaluation using strong LLM judges \cite{chiang2024chatbot}, and have been influential in tracking model quality trends \cite{zheng2023judging}. More fine-grained LLM-based evaluation protocols, such as G-Eval, propose rubric-style prompting to increase alignment with human judgements \cite{liu2023your}. In code-specific settings, recent work has also explored prompt-based "judge” formulations for code correctness decisions, illustrating the appeal of LLM judges when unit tests are unavailable or incomplete \cite{tong2024codejudge}. However, a growing literature warns that LLM judges can be systematically biased and vulnerable. For example, evaluators may exhibit self-preference or other forms of comparative bias, favoring outputs that resemble their own generations rather than outputs that best satisfy the underlying task \cite{panickssery2024llm}. More broadly, work on red-teaming language models emphasizes that model behavior can be manipulated through prompt framing and adversarial or leading instructions, which is directly relevant when judgements are elicited via structured prompts that request explanations and fixes \cite{perez2022redteaming}. 
% These concerns motivate treating prompt design not merely as an accuracy booster, but as a decision boundary control that can shift a model toward conservatism (higher false rejection) or permissiveness (higher false acceptance). 

\section{Methodology}
\subsection{Research Question}
To investigate the reliability, bias patterns, and practical implications of LLM-based requirement conformance judgement \emph{without test cases}, we propose the following research questions:
\begin{itemize}
    \item \textbf{RQ1:} Without test cases, to what extent can LLMs reliably assess whether a program conforms to its specification?
    \item \textbf{RQ2:} How does prompt design affect the LLM-based conformance judgement? In particular, does increasing prompt complexity introduce a systematic tradeoff between \emph{false rejection} and \emph{false acceptance}, and how often do decisions flip across prompting modes?
    \item \textbf{RQ3:} What mechanisms drive false acceptance and false rejection? Specifically, (i) are false positives concentrated on particular bug categories (bug\_type / failure\_symptoms), and (ii) do false negatives exhibit recurring patterns in the model's rejection rationales that can be summarized by a unified taxonomy?
    \item \textbf{RQ4:} How reliable are the explanations produced under rationale-required prompts? whether the rationale supports the stated YES/NO verdict, and the rationale aligns with known bug categories and failure symptoms?
    \item \textbf{RQ5:} What factors cause LLMs to incorrectly classify correct code as faulty, and can these misjudgments be effectively mitigated?
\end{itemize}
These research questions cover three key dimensions: understanding the root causes of LLMs' misjudgments, exploring potential mitigation solutions, and evaluating their effectiveness.

\subsection{Datasets and Construction} \label{sec:dataset}
To enable reliability analysis of requirement conformance judgement, we require paired data that contains both correct and buggy implementations. We construct a unified benchmark suite based on three widely-used code evaluation benchmarks: HumanEval \cite{chen2021evaluating}, MBPP \cite{austin2021programsynthesislargelanguage}, and QuixBugs \cite{ye2021comprehensive}. In total, these benchmarks provide over 700 tasks for evaluation; after pairing each task with both a correct and a buggy implementation, the final dataset contains over 1400 instances. For each task, we create two aligned variants sharing the same requirement text:
(1) a \textbf{canonical} (correct) implementation with label $1$, and
(2) a \textbf{buggy} implementation with label $0$.
This paired design ensures that false rejection (FN) and false acceptance (FP) can be computed consistently across tasks, prompts, and models, rather than only measuring recognition of correct code.

\paragraph{HumanEval-X-Bugs (paired HumanEval).}
For HumanEval, we adopt HumanEval-X-Bugs (from HumanEvalPack \cite{muennighoff2024octopackinstructiontuningcode}), which provides both a canonical solution and a corresponding buggy solution for each task. Each record includes the natural-language task description and the function declaration, enabling us to construct complete Python implementations. Importantly, HumanEval-X-Bugs additionally provides structured labels describing the injected fault, including \texttt{bug\_type} and \texttt{failure\_symptoms}.

\paragraph{Buggy-MBPP construction.}
MBPP \cite{austin2021programsynthesislargelanguage} originally provides task descriptions and correct reference implementations. To support false acceptance analysis, we prepare a paired buggy version of MBPP using an existing bug-aware corpus that provides buggy solution prompts. Since some buggy data is not released as a standalone complete function body, we reconstruct full buggy implementations by combining the buggy prefix/body with the remaining part of the canonical solution, while preserving the same function signature and the overall structure. This yields syntactically complete buggy implementations that are aligned with the original requirements. We further normalize the fault labels into a shared schema so that MBPP, HumanEval-X-Bugs, and QuixBugs can be compared under the same categories.

\paragraph{QuixBugs paired construction.}
QuixBugs \cite{ye2021comprehensive} contains algorithmic problems with known defects and corresponding corrected implementations. We use the corrected version as the canonical implementation and the defective version as the buggy implementation, paired under the same requirement text. Because QuixBugs does not always provide unified fine-grained fault labels in the same form as HumanEval-X-Bugs, we annotate each buggy instance with (i) a normalized \texttt{bug\_type} and (ii) a \texttt{failure\_symptoms} label using a standardized labeling protocol, and then map them into the same schema used for the other benchmarks.

\paragraph{Unified schema and label normalization.}
To support consistent aggregation and cross-benchmark analysis, we convert all three datasets into a unified JSON/JSONL schema with fields such as:
\texttt{task\_id}, \texttt{requirement} (or \texttt{text}), \texttt{variant} (canonical/buggy), \texttt{label} (1/0), \texttt{code}, and (for buggy instances) \texttt{bug\_type} and \texttt{failure\_symptoms}.
Across datasets, we normalize \texttt{bug\_type} into a shared set of six coarse categories (e.g., missing logic, excess logic, operator misuse, variable misuse, value misuse, function misuse), and normalize \texttt{failure\_symptoms} into a compact set describing observable failure modes (e.g., incorrect output, runtime error, non-termination). To facilitate reproducibility, we make the curated datasets and scripts publicly available at \url{https://github.com/HollinJ3177/Are-LLMs-Reliable-Code-Reviewers-Systematic-Overcorrection-in-Requirement-Conformance-Judgement}.

\subsection{Experiment Setup}
To address RQ1 and RQ2, we designed a series of experiments to evaluate the performance of different LLMs on code requirement conformance tasks and to analyze the impact of prompt design. Regarding model selection, we evaluated five LLMs, including three mainstream closed-source models and two representative open-source models. The closed-source models are GPT-4o, Claude-4.5-sonnet, and Google Gemini-2.0-flash. These models were chosen as representatives of state-of-the-art industry LLMs, recognized for their strong general purpose capabilities in code understanding and generation tasks. Their superior performance makes them promising candidates for automated code review \cite{rasheed2024aipoweredcodereviewllms,fragiadakis2025evaluatinghumanaicollaborationreview,zhu2025judgelmfinetunedlargelanguage}. However, precisely because of their capabilities, we aimed to examine if they exhibit consistent error patterns under requirement-conformance judgement. Moreover, these models have consistently been among the top choices for code review and code synthesis tasks in previous works \cite{10403378,joel2024surveyllmbasedcodegeneration,weyssow2024codeultrafeedbackllmasajudgedatasetaligning}. To broaden the coverage and improve reproducibility, we further include two open-source instruction-tuned models: \texttt{meta-llama/llama-3.1-8b-instruct} and \texttt{mistralai/mistral-small-3.1-24b-instruct}. This setting enables a more comprehensive comparison between closed-source and open-source LLMs under identical evaluation protocols.

Fig~\ref{fig:method} summarizes our evaluation pipeline. For each benchmark task, we construct an input package consisting of (i) the natural-language task specification and (ii) two code variants: a canonical reference implementation and a buggy implementation with injected faults, accompanied by the benchmark’s test cases. We then instantiate three prompting modes, for every instance, we record the model outputs in a structured JSON format (including verdict, rationale, and fix when available). Finally, we compute outcome statistics by comparing predicted verdicts against the ground-truth label of each variant (canonical vs.\ buggy), yielding the confusion matrix and the derived error measures used in our subsequent analyses.

\subsection{Prompt Design} \label{sec:prompt}
To ensure a fair comparison, we design three prompting approaches under a unified instruction framework, inspired by recent advances in prompt engineering \cite{wei2023chainofthoughtpromptingelicitsreasoning,yao2023reactsynergizingreasoningacting,jin2024graphchainofthoughtaugmentinglarge,araya2025chainsofthoughtslargelanguagemodels,akbar2024hallumeasure}. Each prompt provides the same input information (the natural language requirement and the code implementation) and differs only in the amount of required intermediate outputs.

\paragraph{Direct prompt.}
The \textsc{Direct} prompt asks the model to read the requirement and the provided code implementation, and answer a single question: ``Does the code meet the requirement?'' The model is instructed to respond with a simple binary answer ``Yes'' or ``No'' only. This prompt represents the minimal setting and serves as a baseline.

\paragraph{Direct + Explain prompt.}
The \textsc{Direct+Explain} prompt extends the baseline by additionally requesting a short rationale for the judgement. Concretely, the model is first asked to output a binary verdict (``Yes''/``No''), and then provide a brief explanation describing why the code meets or does not meet the requirement. This setting isolates the effect of requiring explicit justification, which is commonly believed to encourage more careful reasoning.

\paragraph{Full prompt.}
The \textsc{Full} prompt further adopts a three-step structure:
\begin{itemize}
    \item \textbf{Judgment:} Read the natural language requirement and the provided code implementation, then answer the question, ``Does the code meet the requirement?'' with a simple answer ``Yes'' or ``No''.
    \item \textbf{Explanation:} Request the model to provide a rationale for the judgment, such as explaining why it believes the code does not meet the requirements. This covers a detailed analysis of any discrepancies between code logic and requirements.
    \item \textbf{Fix:} If the model judged the code as not meeting the requirement, it was instructed to provide corrected code after the explanation. If the code was deemed correct, this step could be skipped or explicitly noted as unnecessary.
\end{itemize}

\begin{figure}
    \scriptsize
    \centering
    \includegraphics[width=\linewidth]{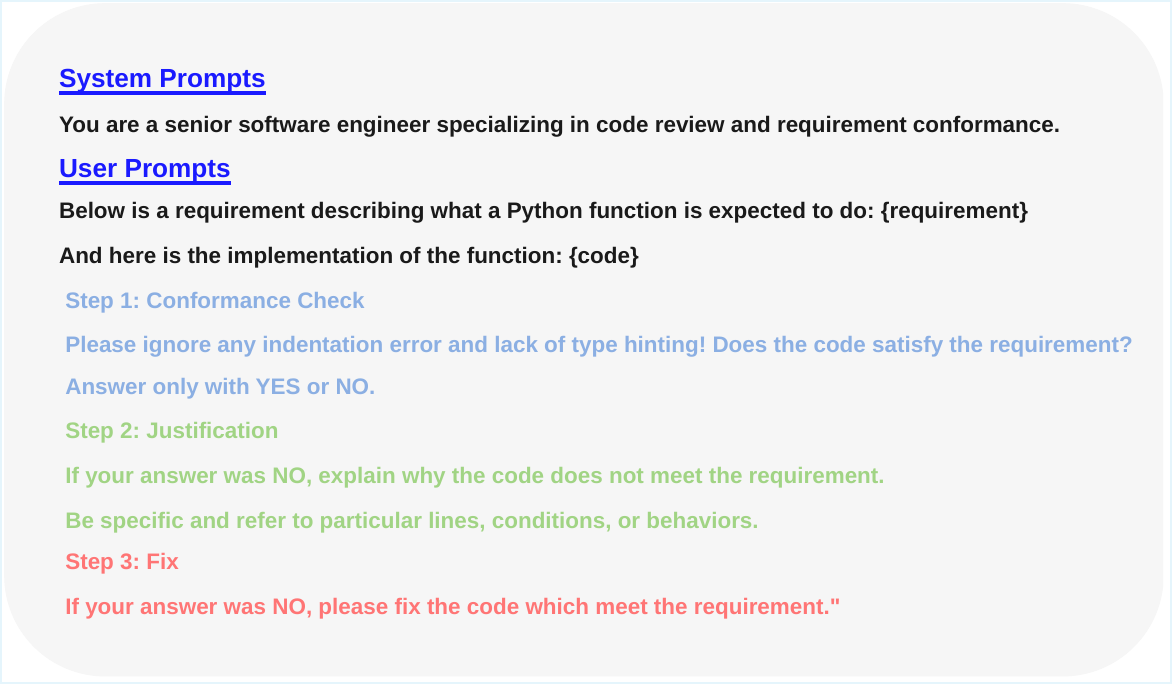}
    \caption{The Full prompt template used in our experiments. It provides the requirement and code, then requests a three-step response.}
    \label{fig:prompt}
\end{figure}
Compared to \textsc{Direct+Explain}, the \textsc{Full} prompt introduces an explicit repair objective, which allows us to examine whether prompting models to ``fix'' issues may inadvertently bias them toward over-correction,  Fig~\ref{fig:prompt} presents examples of the Full prompt setting. 

Multi-stage prompting strategies have been applied for code evaluation task, aiming to enhance the reasoning performance of LLMs. One common approach first instructs the model to perform a step-by-step analysis of the code's functionality, and then asks it to summarize the analysis with a binary decision (correct or not). This approach outperforms simple prompts by encouraging LLMs to reason through both the requirements and the code logic before making a judgment~\cite{tong2024codejudge}. Such findings provide evidence-based support for our design of structured, explanation-driven prompting in code evaluation. Similar two-phase prompting method have also been applied for program repair. For example,~\cite{yin2024thinkrepair,wang2025mcts} prompt LLMs to first generate a chain-of-thought diagnosis of the bug, followed by providing a patch for fixing. Our prompt design is aligned with these approaches, in which the model is initially asked to explain any discrepancies between the code and the requirements, it will use related reasoning to propose a corrected version of the code if the initial answer is negative.

\section{Evaluation Metrics} \label{sec:metrics}
To address our research questions, our evaluation considers two complementary dimensions: (i) the correctness of LLM judgements against ground-truth labels, and (ii) the reliability of the generated rationales in explaining and diagnosing failures. To isolate the impact of prompt complexity, we evaluate three progressively more complex prompting strategies: \textsc{Direct} (judgement only), \textsc{Direct+Explain} (judgement with rationale), and \textsc{Full} (three-step prompt consisting of judgement, explanation, and suggested repair), we report all metrics for each benchmark, each model, and each prompt setting.

\subsection{Confusion matrix outcomes}
With paired canonical/buggy instances (Section~\ref{sec:dataset}), each instance has a binary ground-truth label $y \in \{1,0\}$, where $y=1$ denotes a correct (canonical) implementation and $y=0$ denotes a buggy implementation. Each model outputs a binary verdict $\hat{y} \in \{1,0\}$ parsed from ``Yes/No'' (``Yes'' $\rightarrow 1$, ``No'' $\rightarrow 0$), we compute the following standard outcomes:
\begin{align}
\text{TP} &= \{y=1 \wedge \hat{y}=1\}, \quad
\text{FN} = \{y=1 \wedge \hat{y}=0\}, \nonumber\\
\text{FP} &= \{y=0 \wedge \hat{y}=1\}, \quad
\text{TN} = \{y=0 \wedge \hat{y}=0\}. \nonumber
\end{align}
Intuitively, FN captures \textit{false rejection} (over-correction bias: rejecting correct code), while FP captures \textit{false acceptance} (unsafe acceptance: passing buggy code).
To explicitly characterize these two-sided biases, we primarily report the false negative rate measures the over-correction tendency on correct implementations (rejecting correct code) and false positive rate measures unsafe acceptance on buggy implementations (accepting buggy code), both derived from the confusion matrix. In subsequent analyses, we characterize model behavior primarily through these confusion-matrix counts together with FNR/FPR under different prompting modes and benchmarks.

\begin{table*}
\scriptsize
\centering
\caption{FN reason taxonomy for normalizing the main rationale patterns in those judgments.}
\begin{tblr}{
  colspec = {l X},
  row{1}  = {font=\bfseries},
  column{1} = {halign=l, valign=t},
  column{2} = {halign=l, valign=t},
  hline{1,Z} = {1pt},
  hline{2}   = {0.6pt},
}
Category & Explanation \\
Misread Spec &
The model misreads or overlooks key requirement semantics (e.g., interpreting constraint A as constraint B). \\
Added Requirement &
The model introduces constraints not stated in the requirement and rejects the code for violating these hallucinated requirements. \\
Overthink Edge &
The model overemphasizes extreme or unspecified edge cases, despite the requirement not mandating such handling (or the code already covering it). \\
Assumed Type &
The model assumes stricter (or different) input types/formats and rejects the implementation based on those assumptions. \\
Imagined Runtime &
The model speculates about runtime errors (e.g., index error, \texttt{None} handling, exceptions) without concrete evidence from the given code and requirement. \\
Performance Nitpick &
The model treats efficiency/complexity as a hard requirement even when not mentioned, and rejects correct code due to perceived inefficiency. \\
Readability Nitpick &
The model rejects based on style/readability/best-practice concerns rather than functional conformance. \\
Precision Error &
The model over-concerns floating-point precision/rounding and rejects correct implementations without requirement support. \\
Boundary Error &
The model claims off-by-one or boundary-condition errors (e.g., $<$ vs $\le$) where the implementation is in fact correct. \\
Logic Error &
The model broadly claims incorrect algorithm logic or missing steps, often without a falsifiable counterexample. \\
Vague Description &
The rationale is vague or unsupported, lacking concrete evidence, requirement grounding, or specific failure scenarios. \\
Other &
A catch-all category for unclassifiable patterns. \\
\end{tblr}
\label{tab:fn_taxonomy}
\end{table*}

\subsection{Rejection reason taxonomy} \label{sec:fn_taxonomy}
A key challenge in analyzing false negatives is that, unlike false positives on buggy implementations, FN cases have no ground-truth defect: the implementation is correct (label $1$), yet the LLM rejects it (predicts ``No''). Therefore, the provided \texttt{bug\_type} labels (which describe injected faults in buggy code) are not applicable to FN analysis. Instead, FN cases are characterized by the model's claimed rejection reasons in its rationale, which often reflect systematic over-correction patterns such as requirement hallucination, overemphasis on edge cases, or unsupported assertions. To systematically study these rejection rationales across benchmarks, prompting modes, and model families, we introduce a unified reason taxonomy. Table~\ref{tab:fn_taxonomy} summarizes the taxonomy categories used in our study, each FN instance under rationale enabled prompts is mapped to exactly one primary category based on the main reason expressed in the raw rationale, this normalization enables quantitative aggregation and supports further analysis.

The taxonomy serves three purposes in our study. First, it enables us to quantify which rejection patterns dominate FN cases under different prompts (e.g., whether requiring a ``Fix'' step increases hallucinated requirements or edge-case overconcern). Second, it supports cross-model comparisons to identify whether open-source and closed-source models exhibit different over-correction signatures. Third, it provides a structured lens for qualitative case studies and for designing bias mitigation explorations, by targeting the most frequent rejection patterns rather than treating all FN cases as homogeneous.

\subsection{A1 Self-consistency Evaluation} \label{sec:a1}

While LLMs can output a binary verdict (YES/NO), many prompting approaches additionally request a rationale. However, in practice, the rationale may not faithfully support the final verdict \cite{turpin2023language,paul2024making}. To quantify this phenomenon, we introduce \textbf{self-consistency}, which measures whether an LLM's written rationale aligns with its own stated judgement. A1 is evaluated only for prompting modes that explicitly elicit rationales, i.e., \textsc{Direct+Explain} and \textsc{Full}. For each evaluated instance, we extract:
(1) a normalized verdict $\hat{y}\in\{\text{YES},\text{NO}\}$ from the model output, and
(2) the raw rationale text (the explanation section in the output).

\paragraph{Evaluator and decision criteria.}
We use GPT-4o as an external evaluator to assess whether the rationale supports the stated verdict, the evaluator is instructed \emph{not} to judge the code correctness; instead, it only checks the internal alignment between the verdict and the rationale. For each (verdict, rationale) pair, the evaluator assigns one of three labels:
\texttt{consistent}, \texttt{contradiction}, or \texttt{unclear}.
We additionally record a contradiction type when \texttt{contradiction} is detected:
\texttt{NO\_but\_positive} (verdict is NO but rationale argues the code is correct or compliant),
\texttt{YES\_but\_negative} (verdict is YES but rationale argues the code is incorrect or non-compliant), we also ask the model to output their confidence score and short evidence text to support auditing and qualitative inspection.

% We summarize A1 using contradiction rates conditioned on the verdict, which allows us to capture asymmetric failure modes (e.g., ``NO'' decisions that are justified with positive arguments):
% \begin{equation}
% \label{eq:a1_contra_no}
% \text{ContradictionRate}_{\text{NO}} = 
% \frac{\#\{\texttt{contradiction} \wedge \hat{y}=\text{NO}\}}
% {\#\{\hat{y}=\text{NO}\ \wedge\ \text{A1-scored}\}},
% \end{equation}
% \begin{equation}
% \label{eq:a1_contra_yes}
% \text{ContradictionRate}_{\text{YES}} = 
% \frac{\#\{\texttt{contradiction} \wedge \hat{y}=\text{YES}\}}
% {\#\{\hat{y}=\text{YES}\ \wedge\ \text{A1-scored}\}}.
% \end{equation}
% We report these rates per model, per benchmark, and per prompting mode. Outputs without a parseable verdict or without a rationale are marked as ``skipped'' and excluded from A1 rate computation.

\subsection{A2 Fault-awareness Evaluation} \label{sec:a2}
Beyond internal consistency, a stronger requirement for trustworthy explanations is \emph{fault-awareness}: when the implementation is buggy, does the rationale meaningfully describe the underlying defect and its observable symptoms \cite{turpin2023language}? To answer this question, we introduce \textbf{A2 fault-awareness}, which evaluates whether an LLM rationale aligns with the ground-truth fault labels provided (or normalized) in our paired buggy datasets. A2 is evaluated only on buggy instances (label $0$) where ground-truth fault information is available, including a normalized \texttt{bug\_type} and \texttt{failure\_symptoms} (Section~\ref{sec:dataset}), for each buggy instance, the A2 evaluator receives:
(1) the model's verdict $\hat{y}$,
(2) the raw rationale text,
(3) the ground-truth \texttt{bug\_type}, and
(4) the ground-truth \texttt{failure\_symptoms}. We again use GPT-4o as an external evaluator, instructed to compare the rationale text against the provided ground-truth labels. The evaluator does not decide whether the verdict is correct, it assesses whether the explanation content reflects fault-aware reasoning. For each buggy instance, A2 outputs two alignment judgements:
one for \texttt{bug\_type} and one for \texttt{failure\_symptoms}, each in:
\texttt{match}, \texttt{mismatch}, \texttt{not\_mentioned}, or \texttt{unclear}.
\texttt{not\_mentioned} indicates the rationale does not discuss the relevant fault dimension at all (e.g., only restating the requirement or giving generic comments), while \texttt{unclear} indicates that the rationale mentions a potential issue but the description is too ambiguous to reliably map to the ground-truth label.

\section{Results}
\subsection{Judgement Performance and Bias}\label{sec:result_bias}
Table~\ref{tab:result} reports the false positive rate (FPR) and false negative rate (FNR) of five models across three benchmarks (HumanEval, MBPP, and QuixBugs) under three prompting approaches (\textsc{Direct}, \textsc{Direct+Explain}, and \textsc{Full}). Lower FNR indicates fewer false rejections of correct implementations (i.e., weaker over-correction bias), while lower FPR indicates fewer false acceptances of buggy implementations (i.e., less unsafe acceptance). 

Among the three closed-source models, GPT-4o exhibits the most pronounced over-correction pattern as prompt complexity increases. Under \textsc{Direct}, GPT-4o achieves a relatively low FNR in HumanEval (26.2\%) and MBPP (35.9\%), but once explanations and repairs are required, the FNR increases sharply to 73.2\% in HumanEval and 87.9\% in MBPP. Meanwhile, its FPR drops to nearly zero in several settings (e.g., 0.00\% on HumanEval and MBPP under \textsc{Direct+Explain}), suggesting that the model becomes substantially more conservative, rejecting correct code more often while rarely accepting buggy code.
\begin{table*}
\footnotesize
\centering
\caption{False-positive rate (FPR, \%) and false-negative rate (FNR, \%) of five LLMs on three benchmarks (HumanEval, MBPP, and QuixBugs) under three prompting approaches (\textsc{Direct}, \textsc{Direct+Explain}, and \textsc{Full}). FNR reflects over-correction (rejecting correct implementations), while FPR reflects unsafe acceptance (accepting buggy implementations). Green highlights relatively lower error rates within each model benchmark block, whereas red highlights relatively higher error rates.}
\label{tab:result}
\begin{tblr}{
  column{1} = {c},
  column{2} = {c},
  column{3} = {c},
  column{6} = {c},
  column{9} = {c},
  cell{1}{1} = {r=2}{},
  cell{1}{2} = {r=2}{},
  cell{1}{3} = {c=2}{},
  cell{1}{6} = {c=2}{},
  cell{1}{9} = {c=2}{},
  cell{3}{1} = {r=3}{},
  cell{3}{4} = {c},
  cell{3}{7} = {c},
  cell{3}{10} = {c},
  cell{4}{4} = {c},
  cell{4}{6} = {fg=green},
  cell{4}{7} = {c},
  cell{4}{10} = {c},
  cell{5}{3} = {fg=green},
  cell{5}{4} = {fg=red},
  cell{5}{6} = {c},
  cell{5}{7} = {fg=red},
  cell{5}{9} = {fg=green},
  cell{5}{10} = {fg=red},
  cell{6}{1} = {r=3}{},
  cell{6}{4} = {c},
  cell{6}{6} = {c},
  cell{6}{7} = {c},
  cell{6}{9} = {c},
  cell{6}{10} = {c},
  cell{7}{4} = {c},
  cell{7}{7} = {c},
  cell{7}{10} = {c},
  cell{8}{3} = {fg=green},
  cell{8}{4} = {fg=red},
  cell{8}{6} = {fg=green},
  cell{8}{7} = {fg=red},
  cell{8}{8} = {fg=red},
  cell{8}{9} = {fg=green},
  cell{8}{10} = {fg=red},
  cell{9}{1} = {r=3}{},
  cell{9}{4} = {c},
  cell{9}{7} = {c},
  cell{9}{10} = {c},
  cell{10}{4} = {c},
  cell{10}{7} = {c},
  cell{10}{10} = {c},
  cell{11}{3} = {fg=green},
  cell{11}{4} = {fg=red},
  cell{11}{6} = {fg=green},
  cell{11}{7} = {fg=red},
  cell{11}{9} = {fg=green},
  cell{11}{10} = {fg=red},
  cell{12}{1} = {r=3}{},
  cell{12}{4} = {c},
  cell{12}{7} = {c},
  cell{12}{10} = {c},
  cell{13}{4} = {fg=red},
  cell{13}{6} = {fg=green},
  cell{13}{7} = {fg=red},
  cell{13}{9} = {fg=green},
  cell{13}{10} = {fg=red},
  cell{14}{3} = {fg=green},
  cell{14}{4} = {c},
  cell{14}{6} = {c},
  cell{14}{7} = {c},
  cell{14}{9} = {c},
  cell{14}{10} = {c},
  cell{15}{1} = {r=3}{},
  cell{15}{4} = {c},
  cell{15}{7} = {c},
  cell{15}{10} = {c},
  cell{16}{4} = {c},
  cell{16}{7} = {c},
  cell{16}{10} = {c},
  cell{17}{3} = {fg=green},
  cell{17}{4} = {fg=red},
  cell{17}{6} = {fg=green},
  cell{17}{7} = {fg=red},
  cell{17}{9} = {fg=green},
  cell{17}{10} = {fg=red},
  hline{1,3,6,9,12,15,18} = {-}{},
  hline{2} = {3-4,6-7,9-10}{},
  hline{5,8,11,14,17} = {2-10}{dashed},
}
\textbf{Model}        & \textbf{Approch} & \textbf{HumanEval} &     &  & \textbf{MBPP} &     &  & \textbf{Quixbugs} &     \\
                      &                  & FPR                & FNR &  & FPR           & FNR &  & FPR               & FNR \\
GPT-4o                & Direct           & 2.44               & 26.2   &  & 3.70          & 35.9   &  & 10.9              & 35.0   \\
                      & Direct + Explain & 0.00               & 58.5   &  & 0.00          & 74.1   &  & 5.00              & 45.0   \\
                      & Full             & 0.00~              &~73.2   &  & 0.20          & 87.9   &  & 5.00              & 60.0   \\
Gemini-2.0-flash      & Direct           & 8.54                & 25.6   &  & 10.3          & 34.7   &  & 22.5              & 25.0   \\
                      & Direct + Explain & 7.32               & 23.2   &  & 11.1          & 35.1   &  & 22.5              & 22.5   \\
                      & Full             & 5.49               &~34.1   &  & 7.69          & 39.6   &  & 17.5              & 32.5   \\
Claude-4-5-sonnet     & Direct           & 2.44               & 26.2   &  & 6.57          & 58.5   &  & 5.00              & 40.0   \\
                      & Direct + Explain & 1.21               & 34.1   &  & 6.94         & 55.7   &  & 2.50              & 40.0   \\
                      & Full             & 0.61               &~36.0   &  & 5.44          & 62.3   &  & 2.50              & 50.0   \\
Llama-3.1-8B          & Direct           & 17.1                  & 57.3   &  & 3.56             & 74.7   &  & 27.5                 & 52.5   \\
                      & Direct + Explain & 6.71                  &~86.6   &  & 0.38             & 91.9   &  & 5.00                 & 87.5   \\
                      & Full             & 6.10                  & 84.1   &  & 1.88             & 88.2   &  & 30.0                 & 77.5   \\
Mistral-Small-3.1-24B & Direct           & 6.71                  & 35.9   &  & 5.25             & 60.9   &  & 40.0                 & 40.0   \\
                      & Direct + Explain & 14.6                  & 31.1   &  & 7.13             & 47.8   &  & 40.0                 & 32.5   \\
                      & Full             & 4.88                  &~48.8   &  & 4.31             & 74.3   &  & 27.5                 & 62.5   
\end{tblr}
\end{table*}

Gemini-2.0-flash shows a comparatively different bias profile: its FNR remains moderate (roughly in the 22.5\%-39.6\% range across benchmarks and prompts), but its FPR is consistently higher than GPT-4o and Claude, especially on QuixBugs (22.5\% under \textsc{Direct} and \textsc{Direct+Explain}). This indicates that Gemini is less over-correcting than GPT-4o under richer prompts, yet more prone to false acceptance on buggy implementations. Claude-4.5 generally achieves low FPR (often $\leq$ 6-7\%) while keeping FNR at a moderate-to-high level depending on the benchmark. Notably, on HumanEval, Claude-4.5 maintains relatively low FNR under \textsc{Direct} (26.2\%) but increases under \textsc{Direct+Explain}/\textsc{Full} (36.0\%). On MBPP and QuixBugs, FNR remains substantially higher (e.g., 58.5\% on MBPP under \textsc{Direct}, increasing to 62.3\% under \textsc{Full}), suggesting that for some benchmarks Claude-4.5 also tends toward over-correction, albeit less extremely than GPT-4o.

\begin{rqanswer}{RQ1}
Across three benchmarks, LLMs are not reliably calibrated for requirement-conformance judgement without executing tests. Table~\ref{tab:result} shows substantial false rejection of correct code (high FNR), even under the minimal Direct prompt, and the issue becomes severe under more elaborate prompts. While false acceptance is often lower than FNR, it is non-negligible for some model-benchmark pairs (e.g., higher FPR on QuixBugs for several models), indicating that correctness judgement from specification alone is error-prone and model-dependent.
\end{rqanswer}

\paragraph{Open-source models.}
The two open-source models present higher error rates and stronger sensitivity to prompting. Llama-3.1-8B suffers from extremely high FNR across all benchmarks, indicating severe over-correction: for instance, on MBPP, FNR exceeds 74.7\% even under \textsc{Direct} and rises above 88\% under rationale-enabled prompts. QuixBugs also shows high FNR (52.5\%--87.5\%), implying that the model rejects a large fraction of correct solutions when asked to perform requirement-conformance judgement. Mistral-Small-3.1-24B shows a different failure profile: while its FNR is moderately high across benchmarks, its FPR can be particularly large on QuixBugs (40.0\% under \textsc{Direct} and \textsc{Direct+Explain}). This suggests that, for certain bug patterns in QuixBugs, the model frequently accepts buggy code as conformant, highlighting a risk of unsafe acceptance.

\subsection{Error tradeoff} \label{sec:result_tradeoff}
A key observation from Table~\ref{tab:result} is that prompt design does not uniformly improve judgement quality. Instead, increasing prompt complexity often \emph{redistributes} errors between FPR and FNR, forming a clear tradeoff for several models. Intuitively, prompts that ask the model to provide detailed explanations and propose fixes can encourage model's reasoning, which may reduce false acceptance but simultaneously amplify over-correction bias. For GPT-4o, moving from \textsc{Direct} to \textsc{Direct+Explain} and \textsc{Full} generally decreases FPR while dramatically increasing FNR. The trend is consistent across all three benchmarks, this indicates a strong prompt-induced over-correction effect: richer prompts substantially increase rejection of correct implementations, even though they may also curb false acceptance on buggy ones. 

Claude exhibits a qualitatively similar but milder tradeoff. On HumanEval, FPR decreases monotonically as prompts become more complex (2.44\% to 0.61\%), whereas FNR increases (26.2\% to 36.0\%). On QuixBugs, the same shift appears (FPR: 5.00\% to 2.50\%; FNR: 40.0\% to 50.0\%). These results indicate that increasing prompt complexity (e.g., explanation- and CoT-style prompting) can indeed elicit deeper auditing behavior, making LLMs more likely to surface potential issues in the code. However, this comes with a clear trade-off: the decision boundary shifts toward conservatism, \textbf{reducing false acceptance but amplifying over-correction bias and thus increasing false rejection.} Gemini-2.0-flash shows a weaker strictness shift: \textsc{Full} reduces FPR across benchmarks (e.g., on HumanEval 8.54\% to 5.49\%), while FNR increases only moderately from 25.6\% to 34.1\%.

The open-source models show the strongest non-monotonicity and instability. For Llama-3.1-8B, \textsc{Direct+Explain} sharply increases FNR on all benchmarks (e.g., HumanEval 57.3\% to 86.6\%; MBPP 74.7\% to 91.9\%), while FPR may decrease substantially in some settings (e.g., QuixBugs 27.5\% to 5.00\%). However, adding the repair step in \textsc{Full} does not consistently improve this trade-off (e.g., on QuixBugs, FPR rebounds to 30.0\% while FNR remains high at 77.5\%). For Mistral-Small-3.1-24B, \textsc{Direct+Explain} can even move in the opposite direction on HumanEval by increasing FPR (6.71\% to 14.6\%) while reducing FNR (35.9\% to 31.1\%).

\begin{figure}
    \scriptsize
    \centering
    \includegraphics[width=\linewidth]{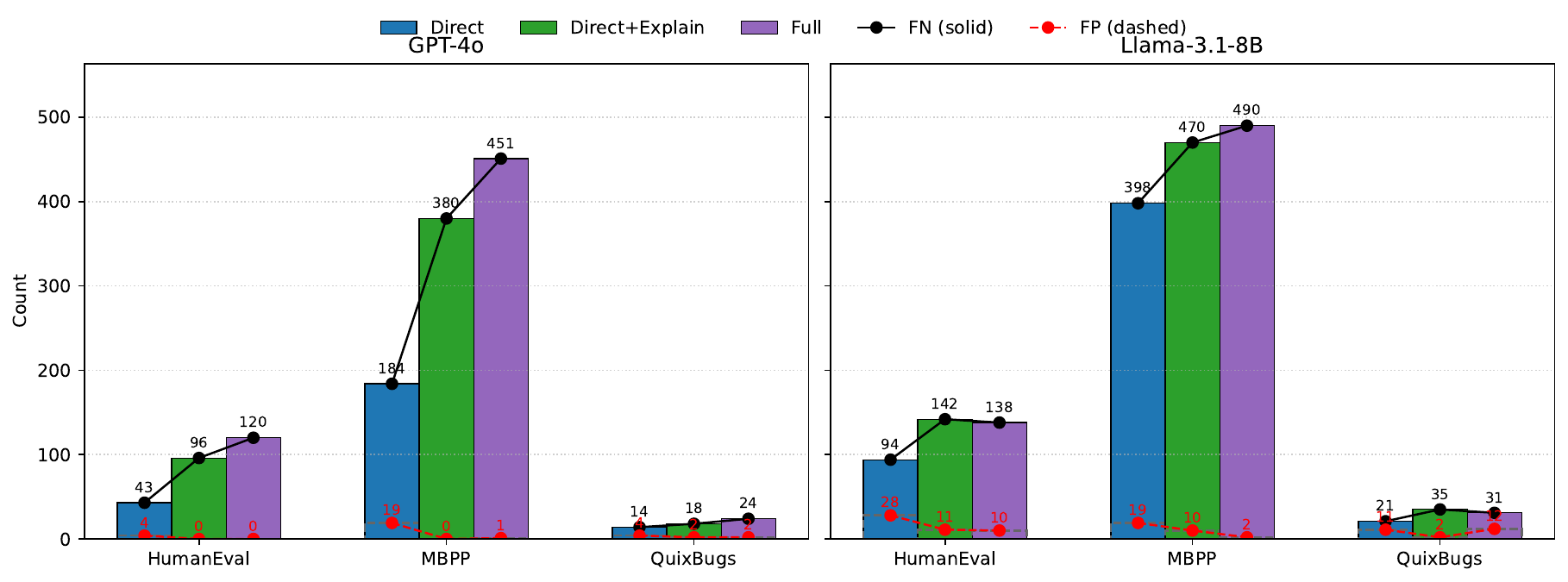}
    \caption{Absolute FN and FP counts across prompt settings for GPT-4o and Llama-3.1-8B.}
    \label{fig:inverse}
\end{figure}

\paragraph{Implications.}
The tradeoff in Table~\ref{tab:result} is not merely a change in percentages, it translates into substantial shifts in the \emph{absolute} number of decisions that developers would have to act on. Fig~\ref{fig:inverse} makes this operational impact explicit. For GPT-4o, richer prompts largely eliminate false acceptance (FP) in some settings (e.g., MBPP FP drops from 19 under \textsc{Direct} to 0 under \textsc{Direct+Explain}, staying near-zero under \textsc{Full}), but the same change simultaneously \emph{more than doubles} false rejection of correct code (FN increases from 184 to 451 on MBPP). A similar pattern appears on HumanEval (FN 43 to 120 while FP 4 to 0). This implies that, in practice, explanation and repair-oriented prompts reduce the small tail of unsafe acceptances, but at the cost of generating a much larger volume of unnecessary rejections and follow-up work, thus increasing the verification burden during the software development.

\begin{rqanswer}{RQ2}
Prompt design systematically reshapes the error profile rather than uniformly improving judgement quality. As prompts become more detailed, several models shift toward conservative rejection: FPR decreases but FNR increases sharply. This tradeoff is operationally large in absolute terms (Fig.~\ref{fig:inverse}), implying frequent decision flips across prompting modes and a substantial increase in unnecessary rejections and follow-up work when explanations or repairs are required.
\end{rqanswer}

For the open-source model (Llama-3.1-8B), Fig~\ref{fig:inverse} suggests an even harsher operational profile: FN dominates across all modes (e.g., MBPP FN 398 to 490), while FP is comparatively smaller and unstable across prompts (MBPP FP 19 to 2). This implies that prompt design should be treated as a \emph{bias control mechanism} rather than a guaranteed accuracy booster: increasing prompt complexity can easily shift the model toward conservative rejection, even when it marginally improves safety against accepting buggy code. Second, the “best” prompt is inherently \emph{cost-sensitive}: scenarios where false acceptance is unacceptable may intentionally adopt stricter prompts and tolerate more FN, while scenarios focused on developer throughput and avoiding unnecessary rework should explicitly control FN and avoid prompts that systematically inflate overcorrection.

\subsection{Sensitivity to bug types} \label{sec:fp_bugtype}
FP analysis is grounded on buggy instances where the injected defect labels (\textit{bug\_type} and \textit{failure\_symptoms}) are available. However, to interpret why models become overly strict (high FNR) or overly permissive (high FPR) under different prompts, it is equally important to examine the \emph{perceived} fault types that models cite in their rationales. In particular, FN cases have no ground-truth defect, so the “bug types” appearing in FN rationales represent the model’s claimed reasons for rejection.
\begin{figure}
    \centering
    \includegraphics[width=\linewidth]{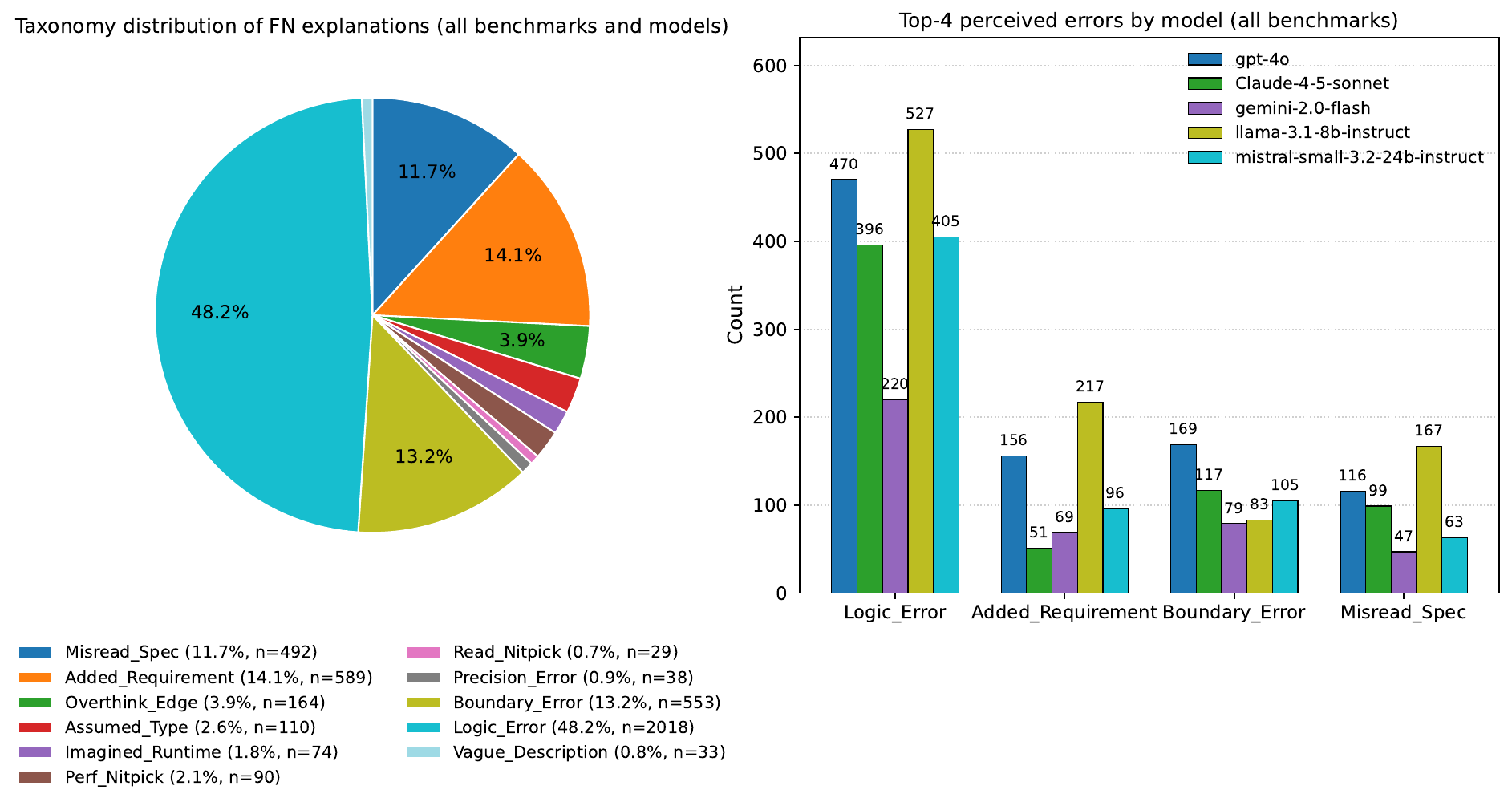}
    \caption{Distribution of perceived fault types in FN rationales (left) and top-4 perceived errors by model (right). Categories include Misread\_Spec (spec misinterpretation), Added\_Requirement (unstated constraints), Overthink\_Edge (boundary overconcern), Assumed\_Type (format assumption), Imagined\_Runtime (unsupported runtime speculation), Perf\_Nitpick (performance critique), Read\_Nitpick (style critique), Precision\_Error (numeric precision), Logic\_Error (algorithmic flaw claim), and Vague\_Description (vague unsupported reasoning).}
    \label{fig:fn_taxonomy}
\end{figure}

As shown in Fig~\ref{fig:fn_taxonomy} left side, nearly half of FN rationales are attributed to \textit{Logic Error} (\textbf{48.2\%}, $n=2018$ out of $4190$), where the model broadly claims the algorithm is wrong or “missing steps,” often without a falsifiable counterexample. The next three categories are \textit{Added Requirement} (\textbf{14.1\%}, $n=589$), \textit{Boundary Error} (\textbf{13.2\%}, $n=553$), and \textit{Misread Spec} (\textbf{11.7\%}, $n=492$). Together, these four categories account for \textbf{87.2\%} of all FN explanations, indicating that over-correction is dominated by semantic failure modes rather than superficial critique. In contrast, the quality related categories are comparatively rare: Perf Nitpick (Performance Complexity Nitpick), Read Nitpick (Style Readability Nitpick), and Precision Error (Numeric Precision) jointly contribute a small fraction of FN explanations. This suggests that the dominant over-correction behavior is not simply the model enforcing style, but rather the model functional mis-calibrating by constructing plausible but unsupported failure bahaviors.

Fig~\ref{fig:fn_taxonomy} (right) further breaks down the top-4 perceived errors by model, we observed that the same four categories (\textit{Logic Error}, \textit{Added Requirement}, \textit{Boundary Error}, \textit{Misread Specification}) dominate across all five models, suggesting these are systematic failure modes. For example, Logic Error is the top perceived error for every model (GPT-4o: $470$; Claude: $396$; Gemini: $220$; Llama: $527$; Mistral: $405$). Second, open-source and closed-source models exhibit different “over-correction signatures.” Llama-3.1-8B shows particularly high counts in Logic Error ($527$) and Added Requirement ($217$), implying a stronger tendency to (i) claim algorithmic flaws, and (ii) introduce extra unstated constraints when rejecting correct code. GPT-4o, while also frequently producing Logic Error rationales, shows a comparatively strong presence of Boundary Error claims ($169$), consistent with a common pattern where the model frames correctness around boundary-case reasoning (e.g., $<$ vs.\ $\le$) even when the implementation is correct. Gemini-2.0-flash has the lowest counts among the five in these categories (e.g., Misread Spec $47$), aligning with its comparatively weaker strictness shift observed in the error-tradeoff analysis.

\begin{rqanswer}{RQ3}
We find recurring and highly concentrated mechanisms behind false rejections. By normalizing FN rationales into a unified taxonomy, four dominant patterns account for 87.2\% of FN explanations: \textit{Logic Error} (48.2\%), \textit{Added Requirement} (14.1\%), \textit{Boundary Error} (13.2\%), and \textit{Misread Spec} (11.7\%) (Fig.~\ref{fig:fn_taxonomy}). These results indicate that over-correction is primarily driven by unverified claims and requirement hallucination (inventing unstated constraints), rather than superficial style critique. For buggy implementations, model performance varies strongly by benchmark and model family (Table~\ref{tab:result}), suggesting that false acceptance is sensitive to the bug patterns and oracle strength of each benchmark (e.g., QuixBugs exhibits higher and more unstable FPR for several models).
\end{rqanswer}

These findings suggest that both FP and FN analyses \textbf{should not} treat errors as homogeneous. Instead, the most actionable leverage comes from targeting a small set of recurrent “fault narratives.” Specifically, the dominance of Logic Error and Added Requirement indicates that a large portion of rejections are driven by unverified claims. This motivates mitigation strategies that explicitly force evidence grounding, for example, \textbf{requiring the rationale to cite the exact requirement clause being violated and to provide a concrete counterexample input/output trace and that discourage unconstrained constraint invention}. Likewise, the prevalence of Boundary Error suggests that models may be over-applying “bug priors” (boundary mistakes are common in real code) as if they were proof of non-conformance, motivating prompts that separate “possible risk” from “confirmed violation.”

\section{Explanation Reliability}
\subsection{A1 Rationale self-consistency}
LLM-based code review tools are typically consumed through explanations rather than binary verdicts alone, 
therefore, beyond judgment correctness, we evaluate whether the rationale of a model is internally consistent with its own verdict. In A1, we focus on rationale enabled prompting modes (\textsc{Direct+Explain} and \textsc{Full}) and use a GPT-4o evaluator to label each output as consistent, contradiction, or unclear. 
We treat contradiction and unclear as \emph{inconsistent rationales}, since both reflect unreliable explanation quality: contradictions explicitly argue against the verdict, while unclear rationales are too vague or mixed to justify the decision.

\begin{figure}
    \centering
    \includegraphics[width=\linewidth]{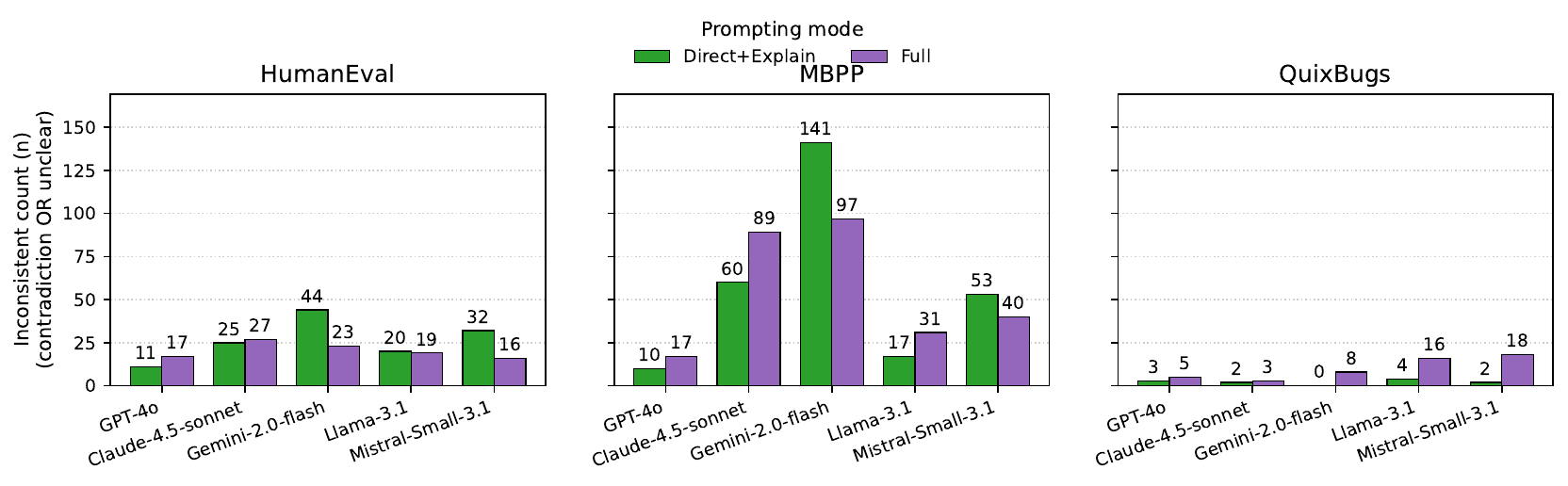}
    \caption{Counts of inconsistent rationales (labeled as contradiction or unclear by a GPT-4o evaluator) under rationale-enabled prompts. Bars compare Direct+Explain vs.\ Full for each model across HumanEval, MBPP, and QuixBugs, report absolute counts.}
    \label{fig:A1-1}
\end{figure}

\paragraph{Inconsistency is non-trivial and benchmark dependent.}
Fig~\ref{fig:A1-1} reports the number of inconsistent rationales across the three benchmarks.
Overall, inconsistency is not a rare corner case: even strong closed-source models can produce a noticeable amount of contradiction or unclear rationales \cite{lanham2023measuring}. For instance, GPT-4o shows consistent increases in inconsistency when moving from \textsc{Direct+Explain} to \textsc{Full} across all benchmarks (HumanEval: $11\rightarrow17$, MBPP: $10\rightarrow17$, QuixBugs: $3\rightarrow5$). Claude-4.5-sonnet exhibits a similar trend, but with a substantially larger jump on MBPP ($60\rightarrow89$), indicating that longer, repair-oriented prompting can destabilize its own justification under more diverse tasks.

Meanwhile, the effect of adding the repair step is not uniform across models. Gemini-2.0-flash becomes more self-consistent on HumanEval and MBPP (HumanEval: $44\rightarrow23$, MBPP: $141\rightarrow97$), but becomes less self-consistent on QuixBugs ($0\rightarrow8$). Open-source models show the clearest dataset sensitivity on QuixBugs: Llama-3.1’s inconsistency rises from $4$ to $16$, and Mistral-Small-3.1 rises from $2$ to $18$ when switching from Direct+Explain to Full. This is consistent with the intuition that the carefully crafted prompt introduces additional degrees of freedom (repair planning and patch justification), which can amplify explanation instability, especially on small bug-centric benchmarks.

% typo on claude 3.5
\begin{figure}
    \centering
    \includegraphics[width=\linewidth]{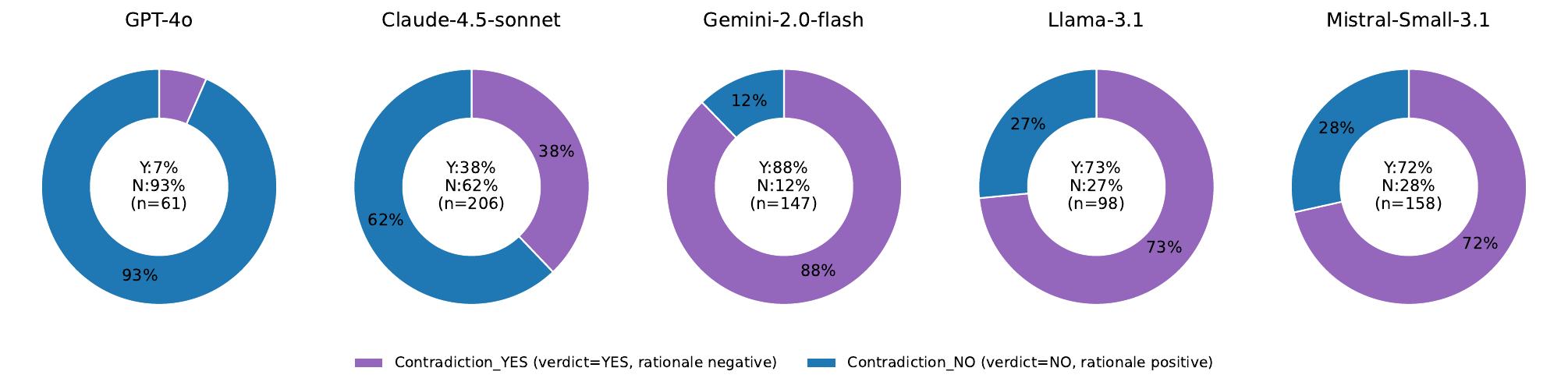}
    \caption{Directional breakdown of contradiction cases by model. Each chart shows the fraction of Contradiction\_NO (verdict=NO but rationale positive) vs.\ Contradiction\_YES (verdict=YES but rationale negative), center labels report percentages and the total number of contradictions ($n$).}
    \label{fig:A1-2}
\end{figure}

\paragraph{Contradictions have model-specific directions.}
Fig~\ref{fig:A1-2} further decomposes contradiction cases into two interpretable types:
(\textit{i}) \textbf{Contradiction\_NO} (verdict=NO, but the rationale is positive), and
(\textit{ii}) \textbf{Contradiction\_YES} (verdict=YES, but the rationale is negative).
We observe a clear directional asymmetry that varies across model families. For GPT-4o, most contradictions are \textbf{Contradiction\_NO} (93\%, $n{=}61$). In other words, when GPT-4o contradicts itself, it usually rejects the code even though its explanation reads as if the code should be accepted. This suggests that under richer prompts, GPT-4o becomes more conservative in its final verdict, but its explanation does not consistently provide a strong or concrete reason for rejection.

In contrast, Gemini-2.0-flash shows the opposite behavior: 88\% of contradictions are \textbf{Contradiction\_YES} ($n{=}147$), it outputs “YES” while simultaneously highlighting potential defects. 
A similar tendency appears in the open-source models (Llama-3.1: 73\% \textbf{Contradiction\_YES}, $n{=}98$; Mistral-Small-3.1: 72\% \textbf{Contradiction\_YES}, $n{=}158$). This suggests a pattern that the model preserves an affirmative verdict but produces a rationale that is not aligned with that verdict, which can be particularly misleading in practice because it blurs the meaning of a “YES” judgement \cite{paul2024making}. Claude-4.5-sonnet is comparatively more balanced, implying that its inconsistencies arise from both conservative and permissive failure modes. These results demonstrate that more detailed prompting does not guarantee explanation reliability: richer prompts may elicit longer rationales, but do not ensure that the rationale coherently supports the model’s own verdict. Moreover, the direction of contradiction is model-dependent: GPT-4o tends to reject with supportive rationales, whereas Gemini and the open-source models tend to accept with fault-claiming rationales, practically, this undermines the common assumption that “adding explanations increases trustworthiness.” 

\subsection{A2 Fault-awareness}
A2 evaluates whether an LLM’s rationale is fault-aware on buggy instances: when the implementation is incorrect, does the explanation correctly describe (i) the bug type (root cause category) and (ii) the failure symptoms (observable consequence)? Importantly, A2 does not judge whether the verdict itself is correct instead, it checks whether the content of the rationale aligns with the ground-truth fault labels. 
% We report A2 under rationale-enabled prompts and focus on cases where the model verdict is \textbf{correct} on buggy instances (\textbf{TN}), since this setting best reflects a realistic “LLM reviewer” scenario: even if the reviewer correctly rejects buggy code, the rationale can still be unhelpful or misleading if it fails to identify the actual defect.
\begin{table*}[t]
\centering
\scriptsize
\setlength{\tabcolsep}{3pt}
\caption{A2 fault-awareness on buggy instances when the model verdict is correct. We report match rates (\%) between the rationale and ground-truth bug type matched and failure symptoms matched.}
\label{tab:A2-1}
\begin{tblr}{
  colspec = {l r r r r r r},
  row{1} = {font=\bfseries},
  cell{1}{2} = {c=2}{}, cell{1}{4} = {c=2}{}, cell{1}{6} = {c=2}{},
  hline{1,Z} = {1pt}, hline{2} = {0.6pt},
}
Model & \textbf{HumanEval} & & \textbf{MBPP} & & \textbf{QuixBugs} & \\
      & BugMatch & SymMatch & BugMatch & SymMatch & BugMatch & SymMatch \\
GPT-4o & 59.1 & 98.2 & 70.8 & 94.7 & 58.3 & 100.0 \\
Claude-4.5 & 65.0 & 97.5 & 75.2 & 93.3 & 57.9 & 92.1 \\
Gemini-2.0 & 52.0 & 98.7 & 67.2 & 95.8 & 38.2 & 91.2 \\
Llama-3.1 & 44.2 & 96.8 & 52.2 & 96.0 & 41.7 & 100.0 \\
Mistral-3.1 & 50.0 & 96.8 & 65.7 & 94.5 & 58.3 & 95.8 \\
\end{tblr}
\end{table*}

\begin{figure}
    \centering
    \includegraphics[width=\linewidth]{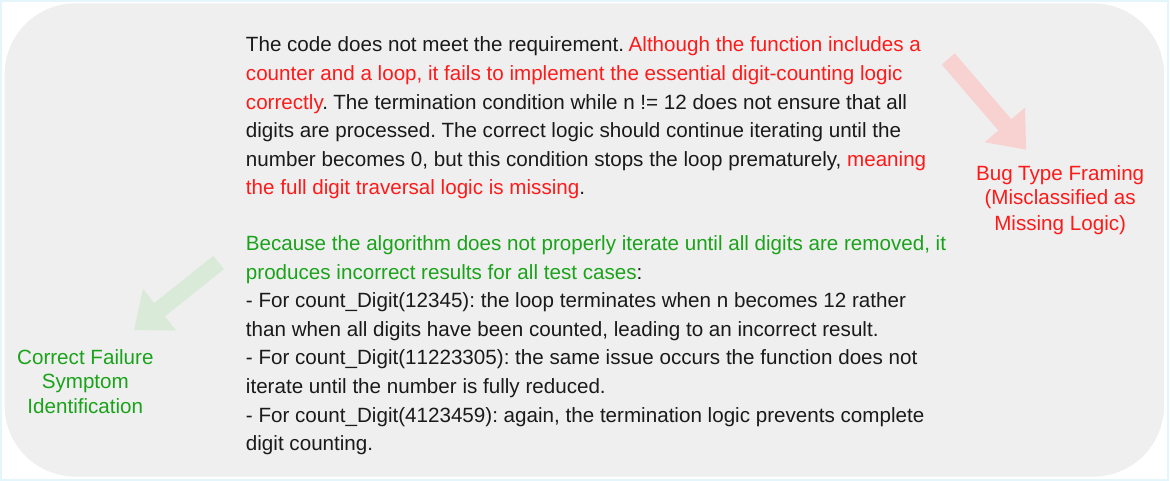}
    \caption{Example of a symptom is correct but bug type is not matched, the model correctly states the failure symptom, but frames the root cause as Missing Logic (highlighted in red) instead of the ground-truth Value Misuse.}
    \label{fig:bug_misclassify}
\end{figure}

\paragraph{Symptom awareness is high, bug-type awareness is substantially lower.}
Table~\ref{tab:A2-1} summarizes A2 match rates under the detailed prompt across three benchmarks.
A consistent pattern emerges across all models and datasets: \textbf{SymptomMatch is near-ceiling}, while \textbf{BugMatch is much lower and varies substantially by benchmark and model}.
For example, GPT-4o achieves SymptomMatch of 98.2\% (HumanEval), 94.7\% (MBPP), and 100.0\% (QuixBugs), but its BugMatch is only 59.1\%, 70.8\%, and 58.3\%, respectively, this gap is also visible for the other models. Open-source models show the same qualitative discrepancy: Llama-3.1 has SymptomMatch $\geq$ 96\% across benchmarks, yet BugMatch is only 44.2\% (HumanEval), 52.2\% (MBPP), and 41.7\% (QuixBugs). Taken together, these results indicate that \textbf{correct rejection (TN) is frequently driven by symptom-level reasoning} (e.g., “this will produce incorrect output” or “this can fail at runtime”), while \textbf{cause-level diagnosis is much less reliable}. Practically, this means that an LLM reviewer can be “right for the inaccurate reason”: it rejects buggy code correctly, but its explanation may still mislead debugging by attributing the failure to an incorrect bug category. Fig~\ref{fig:bug_misclassify} shows a concrete example: the rationale correctly identifies the failure symptom (incorrect output), but mis-classifies the underlying cause by framing it as \textit{Missing Logic} even though the ground-truth defect is \textit{Value Misuse}.

\begin{figure}
    \centering
    \includegraphics[width=\linewidth]{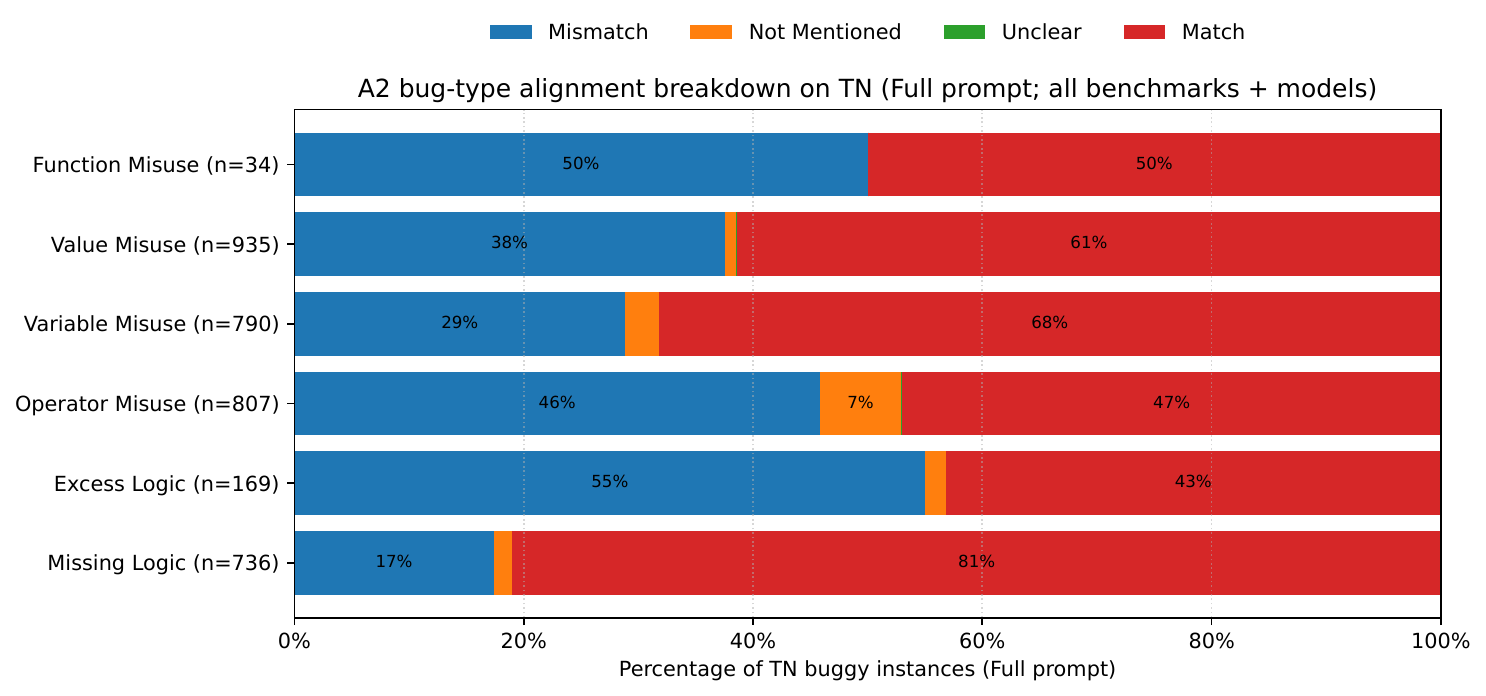}
    \caption{bug-type alignment on TN buggy instances, Bars report the distribution of rationale labels (match, mismatch, not mentioned, unclear) with respect to ground-truth bug types.}
    \label{fig:A2-2}
\end{figure}

\paragraph{Bug-type breakdown.}
To understand why BugMatch is low, we further decompose the bug-type alignment outcomes (match / mismatch / not mentioned / unclear) by ground-truth bug type in Fig~\ref{fig:A2-2} (aggregated over all benchmarks and models under the Full prompt, restricted to TN buggy instances). First, \textbf{bug-type match is highly bug-dependent}. \textit{Missing Logic} is the easiest category: it reaches \textbf{81\% match} with only \textbf{17\% mismatch}, suggesting that when the defect is framed as an omission (e.g., missing a required check or missing a key step) \cite{tong2024codejudge,paul-etal-2024-making}, rationales more often converge to the correct causal description. In contrast, \textit{Excess Logic} and \textit{Operator Misuse} show much weaker alignment: \textit{Excess Logic} has 55\% mismatch, and \textit{Operator Misuse} is nearly split (46\% mismatch47).
\textit{Function Misuse} is also unstable, although it has the smallest support ($n=34$), so conclusions for this category should be treated cautiously. Second, \textbf{low BugMatch is more often caused by misdiagnosis than by silence}. Nevertheless, \textbf{not mentioned} is not negligible for some bug types: \textit{Operator Misuse} shows the highest not-mentioned proportion (\textbf{7\%}), indicating that even in TN cases, a subset of rationales rejects buggy code while failing to explicitly articulate the underlying fault category. This provides direct evidence for our central A2 claim: \textbf{verdict correctness does not guarantee fault-aware explanations}.

The A2 results complement our earlier judgement-level bias analysis: while prompt engineering can shift verdict-level behavior, the prompt engineering does not ensure that “more reasoning” yields more diagnostic explanations. Table~\ref{tab:A2-1} shows that rationales are generally reliable at stating what goes wrong, but Fig~\ref{fig:A2-2} shows that for several bug classes they are much less reliable at stating why it goes wrong, with mismatch being a prevalent failure mode. This mismatch matters for downstream software engineering workflows: a rationale that correctly flags a defect but attributes it to the wrong cause can still induce ineffective or even harmful repair suggestions, and can waste developer time during debugging \cite{tong2024codejudge}. Consequently, A2 suggests that LLM-based code reviewer evaluation should separate decision correctness from explanation fault-awareness, and that mitigation efforts should explicitly target cause-level grounding rather than assuming that longer rationales or repair requests automatically improve diagnosis quality.

\begin{rqanswer}{RQ4}
Explanations elicited by rationale required prompts are not consistently reliable. Under A1, we observe non-trivial rates of internally inconsistent rationales (contradiction/unclear), and the contradiction direction is model-dependent. Under A2, rationales are generally strong at describing symptoms but weaker at diagnosing root causes: SymptomMatch is near-ceiling across benchmarks, whereas BugMatch is substantially lower and varies by model (Table~\ref{tab:A2-1}). Overall, longer explanations do not guarantee faithful justification, and correct verdicts can still be paired with misleading bug diagnoses.
\end{rqanswer}

\section{Bias mitigation exploration}
\subsection{Setup and strategies}
Our earlier results reveal a pronounced over-correction bias under Full prompting: when asked to (i) judge conformance, (ii) justify, and (iii) propose a fix, LLMs often become overly conservative and reject correct implementations, leading to a high false-negative rate, while the false-positive rate is comparatively low. This pattern suggests that increasing prompt complexity (e.g., explanation and repair-oriented instructions) can indeed surface more potential issues, but it also shifts the decision boundary toward stricter rejection, amplifying unnecessary “repair” behaviors. 

\begin{figure}
    \centering
    \includegraphics[width=\linewidth]{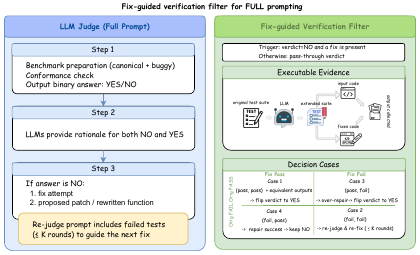}
    \caption{Overview of the Fix-guided Verification Filter. When the judge outputs NO with a proposed fix, we execute both the original and fixed implementations on the benchmark tests and generated spec-constrained tests, and update the final verdict according to four executable cases.}
    \label{fig:bias_miti}
\end{figure}

To mitigate this bias, we introduce a lightweight Fix-guided Verification Filter (Fig.~\ref{fig:bias_miti}) that sits inside the evaluation pipeline (rather than post-hoc editing stored predictions). The key intuition is to treat the model’s proposed fix as executable evidence: if a model says \texttt{NO} and proposes a patch, then the pair (\textit{original code}, \textit{fix code}) forms a natural counterfactual for verification. Instead of trusting the textual rationale, we validate both programs with runtime evidence using (1) the benchmark’s reference test suite and (2) a spec-constrained augmented test suite generated by GPT-4o. This design is inspired by the ``generate-and-validate'' philosophy in test suite based program repair and execution guided verification, where candidate changes are accepted only when they withstand behavioral checks, while also aligning with recent self-refinement or feedback-based LLM frameworks that leverage external signals to correct model behaviors \cite{madaan2023self,shinn2023reflexion}.

% decision matrix rather than decision cases

\paragraph{Filter decision cases.}
Given an input program $c$ and a proposed fix $\hat{c}$ (extracted from the response), the filter executes both against the benchmark test cases $T$ and an augmented test set $\tilde{T}$. The final verdict is determined by four common outcomes:

\begin{itemize}
    \item \textbf{Case 1 (equivalent):} $c$ and $\hat{c}$ both pass, and they are behaviorally consistent (same outputs) on spec-constrained augmented inputs $\tilde{T}$. This indicates the fix is unnecessary, we treat the original decision as over-correction and flip the verdict to YES.
    \item \textbf{Case 2 (both fail):} both $c$ and $\hat{c}$ fail on the augmented tests, suggesting either the augmentation over-extended the spec or the repair is still incorrect. We fall back to the benchmark tests only and allow up to $K$ re-judge rounds, where the next judgment is conditioned on previously failed tests to guide refinement, if the process remains unresolved after $K$ rounds, we keep the original verdict.
    \item \textbf{Case 3 (over-repair):} $c$ passes but $\hat{c}$ fails, meaning the proposed fix introduces regressions, we interpret this as over-repair and flip to YES.
    \item \textbf{Case 4 (repair success):} $c$ fails but $\hat{c}$ passes, indicating the fix corrects genuine nonconformance, we keep the original verdict.
\end{itemize}

\paragraph{Implementation setup.}
The filter is applied only when the judge returns \texttt{NO} and provides a fix. All five models (GPT-4o, Gemini-2.0-flash, Claude-4.5-sonnet, Llama-3.1-8B, and Mistral-Small-3.1-24B) produce the initial verdict and fix, while the augmented test generation step is standardized to GPT-4o to ensure consistent filtering across judge models. Augmented tests are explicitly constrained by the task specification and examples to avoid introducing adversarial requirements, in our experiments we use $K{=}2$ rounds, and cache augmented tests to improve reproducibility and robustness.

\subsection{Result and Analysis}
Table~\ref{tab:mitigation_filter_result} reports FPR/FNR before vs.\ after applying the Fix-guided Verification Filter under Full prompting. Averaged across models, FNR drops from 54.8\% to 16.3\% on HumanEval and 69.0\% to 28.9\% on MBPP, while QuixBugs also improves (51.0\% to 24.0\%). Notably, the gains are strongest for models with extreme baseline conservatism under the Full setup, e.g., Llama-3.1-8B on MBPP (90.81\%$\rightarrow$23.56\%) and GPT-4o on MBPP (88.74\%$\rightarrow$39.96\%), demonstrating that the filter can substantially counteract prompt-induced strictness.

\begin{table*}[t]
\footnotesize
\centering
\caption{False-positive rate (FPR, \%) and false-negative rate (FNR, \%) before vs.\ after applying the Fix-guided Verification Filter under \textsc{Full} prompting. For each \textsc{Full+Filter} row, the FNR cell additionally reports the absolute reduction (percentage points) from the corresponding \textsc{Full} baseline as $\downarrow(\Delta)$.}
\label{tab:mitigation_filter_result}
\begin{tblr}{
  colspec = {c c c c c c c c},
  colsep = 3pt,
  column{1} = {c},
  column{2} = {c},
  column{3} = {c}, column{4} = {c},
  column{5} = {c}, column{6} = {c},
  column{7} = {c}, column{8} = {c},
  cell{1}{1} = {r=2}{},
  cell{1}{2} = {r=2}{},
  cell{1}{3} = {c=2}{},
  cell{1}{5} = {c=2}{},
  cell{1}{7} = {c=2}{},
  cell{3}{1}  = {r=2}{},
  cell{5}{1}  = {r=2}{},
  cell{7}{1}  = {r=2}{},
  cell{9}{1}  = {r=2}{},
  cell{11}{1} = {r=2}{},
  hline{1,3,5,7,9,11,13} = {-}{},
  hline{2} = {3-4,5-6,7-8}{},
  hline{4,6,8,10} = {2-8}{dashed},
}
\textbf{Model} & \textbf{Setting} &
\textbf{HumanEval} & & \textbf{MBPP} & & \textbf{QuixBugs} & \\
& & FPR & FNR & FPR & FNR & FPR & FNR \\

GPT-4o & Full (Before)   & 0.0 & 70.7 & 0.0 & 88.7 & 5.0 & 57.5 \\
      & Full+Filter (After) & 1.8 & 23.2$\downarrow$(\textcolor{green!60!black}{47.6}) & 0.4 & 40.0$\downarrow$(\textcolor{green!60!black}{48.8}) & 7.5 & 27.5$\downarrow$(\textcolor{green!60!black}{30.0}) \\

Gemini-2.0-flash & Full (Before)   & 7.8 & 25.0 & 11.1 & 30.1 & 22.5 & 20.0 \\
                 & Full+Filter (After) & 6.4 & 5.8$\downarrow$(\textcolor{green!60!black}{19.2})  & 10.2 & 10.2$\downarrow$(\textcolor{green!60!black}{19.9}) & 22.5 & 12.5$\downarrow$(\textcolor{green!60!black}{7.5}) \\

Claude-4-5-sonnet & Full (Before)   & 1.9 & 46.0 & 5.3 & 61.2 & 5.0 & 50.0 \\
                  & Full+Filter (After) & 3.1 & 8.7$\downarrow$(\textcolor{green!60!black}{37.3})  & 5.6 & 27.2$\downarrow$(\textcolor{green!60!black}{34.0}) & 7.5 & 22.5$\downarrow$(\textcolor{green!60!black}{27.5}) \\

Llama-3.1-8B & Full (Before)   & 4.9 & 82.3 & 1.5 & 90.8 & 17.5 & 70.0 \\
             & Full+Filter (After) & 6.1 & 27.4$\downarrow$(\textcolor{green!60!black}{54.9}) & 1.7 & 23.6$\downarrow$(\textcolor{green!60!black}{67.3}) & 20.0 & 25.0$\downarrow$(\textcolor{green!60!black}{45.0}) \\

Mistral-Small-3.2-24B & Full (Before)   & 3.7 & 50.0 & 4.3 & 74.1 & 22.5 & 57.5 \\
                      & Full+Filter (After) & 5.5 & 16.5$\downarrow$(\textcolor{green!60!black}{33.5}) & 4.7 & 43.5$\downarrow$(\textcolor{green!60!black}{30.6}) & 25.0 & 32.5$\downarrow$(\textcolor{green!60!black}{25.0}) \\
\end{tblr}
\end{table*}
The observed improvements are aligned with the mechanism of the filter: when an LLM proposes a fix for a correct program, that fix often does not introduce new behavior, by requiring the original and fixed code to be consistent under executable tests (including spec-constrained augmentation), the filter rejects spurious NO decisions and recovers correct YES outcomes. Conceptually, this resembles ``execution as an arbiter'' and ``patch validation'' in test-driven repair, where runtime evidence is used to gate model-driven changes and avoid unnecessary edits. In our setting, the same idea serves as a bias mitigation tool: it converts the model’s own patch into a verifiable hypothesis and uses differential evidence to decide whether the initial rejection was warranted. While the filter is designed primarily to mitigate over-correction bias, FPR can increase slightly after filtering (e.g., GPT-4o on HumanEval: 0.00\%$\rightarrow$1.83\%; several models on QuixBugs: +2.5 points). This is expected: the filter is triggered only on NO+fix cases, and flipping some of these rejections to YES can inadvertently accept a small number of buggy programs that pass limited tests. This effect is most visible on QuixBugs, where the reference test suites are often shallow, reducing the discriminative power of execution-based checks; even spec-constrained augmentation cannot fully compensate for missing oracle coverage. Practically, this suggests the filter is most reliable when the benchmark harness provides reasonable behavioral coverage (as in HumanEval/MBPP), and it motivates future extensions such as stronger augmentation (e.g., metamorphic relations or property-based generators), multi-oracle agreement, or combining execution evidence with lightweight static checks to further limit residual false acceptances. Overall, Table~\ref{tab:mitigation_filter_result} shows that the Fix-guided Verification Filter provides a simple and model-agnostic way to substantially reduce over-correction bias under Full prompting, by grounding verdict revision in executable evidence rather than purely textual rationales.
\begin{rqanswer}{RQ5}
False rejection of correct code is primarily caused by prompt-induced conservatism and rationale-driven overreach: models frequently hallucinate extra constraints, over-emphasize edge cases, or assert vague ``logic errors'' without executable evidence (Fig.~\ref{fig:fn_taxonomy}). To mitigate this over-correction bias, we introduce a Fix-guided Verification Filter that treats the model-proposed fix as an executable counterfactual and validates under benchmark and spec-constrained augmented tests (Fig.~\ref{fig:bias_miti}). The filter substantially reduces FNR across all five models and three benchmarks, with only modest FPR increases in some settings, this suggests misjudgments can be mitigated by grounding decisions in executable evidence rather than relying solely on increasingly elaborate prompts.
\end{rqanswer}

\section{Threats to Validity}
This section discusses limitations and potential confounding factors in our experimental design and the proposed Fix-guided Verification Filter, organized following common validity dimensions.

\subsection{Construct Validity}
A central construct in our study is over-correction (rejecting correct implementations), which is closely related to but not identical to LLMs hallucination \cite{ji2023survey}. Many false rejections are accompanied by confident rationales that introduce unstated constraints or speculative failure scenarios, which resembles requirement hallucination in natural-language generation. However, our current analysis does not explicitly align the taxonomy of FN rationales with established hallucination definitions and measurements (e.g., intrinsic vs.\ extrinsic hallucination) \cite{alansari2025largelanguagemodelshallucination}. Consequently, some of the observed misjudgments may be partially explained by broader hallucination behaviors rather than a unique code-review specific bias. Future work could bridge this gap by jointly evaluating FN rationales with hallucination-oriented metrics and annotator guidelines, and by explicitly modeling whether each rejection claim is supported by the given specification.

\textbf{Operationalization of correctness via benchmark labels and tests.}
We treat canonical solutions as correct (label $=1$) and buggy solutions as incorrect (label $=0$), and compute FNR/FPR accordingly. While this paired design enables controlled comparisons, it assumes that canonical solutions and embedded tests reflect the intended specification. In practice, benchmark tests can be incomplete, and canonical solutions may encode one of several valid behaviors consistent with the natural-language requirement. As a result, a model may be penalized for rejecting a canonical implementation due to an alternative (but still plausible) spec interpretation, or a buggy implementation may slip through weak tests and appear acceptable. This is particularly salient for small suites (e.g., QuixBugs), where limited oracle coverage can blur the boundary between functional non-conformance and under-specification.

\textbf{Prompt-mode construct and interpretation.}
Our three prompting modes (\textsc{Direct}, \textsc{Direct+Explain}, \textsc{Full}) are meant to represent progressively richer reviewer instructions. Still, prompt semantics can implicitly signal different review criteria (e.g., encouraging best-practice critique or performance concerns) even when not required by the task \cite{zhu2023promptrobust}. Therefore, part of the measured ``bias shift'' may reflect a change in the latent evaluation rubric induced by the prompt, rather than a pure reasoning-quality change.

\subsection{Internal Validity}
\textbf{Sensitivity to prompt wording and formatting.}
Results can change with prompt phrasing. Although we use common, widely adopted prompt patterns, minor changes (e.g., stricter instruction to ignore efficiency, different formatting of requirements/code blocks, or alternative delimiters) can alter the model's decision boundary and thus the measured FNR/FPR. This introduces a threat that some effects may be prompt-instance specific, a stronger design would include multiple paraphrased prompts per mode and report variance (or worst/best-case bounds) across prompt templates.

\textbf{Non-determinism and API/runtime variability.}
Even with temperature $=0$, model-serving infrastructure can introduce slight non-determinism (e.g., backend updates, routing, transient failures). Additionally, tool-chain details (Python version, timeouts, process isolation) may affect whether executions terminate or time out, which can influence filter decisions. We partially mitigate this by logging raw outputs, caching augmented tests, and using consistent execution budgets, but residual nondeterminism may still affect edge cases.

\textbf{Verification is only as strong as the tests.}
The Fix-guided Verification Filter relies on the benchmark test suite plus GPT-generated spec-constrained augmentations. If the augmented tests fail to cover the relevant behavioral space, the filter may incorrectly treat two implementations as behaviorally equivalent (Case 1) or accept a buggy implementation that still passes the available tests. This is a key reason why mitigation can reduce FNR substantially while sometimes slightly increasing FPR: flipping some NO to YES decisions can inadvertently pass buggy code under insufficient test coverage, this is an inherent limitation of execution-based validation without a formal oracle.

\textbf{Evaluator coupling and potential judge bias in augmentation.}
Although the initial verdicts and fixes come from five different judge models, the augmentation step is standardized to GPT-4o. This improves consistency between conditions, but introduces dependency on a single model's prior testing. If GPT-4o systematically under-generates certain edge cases, the filter may overestimate equivalence. Conversely, if it over-generates constraints, the filter may become overly strict. A more robust alternative would triangulate augmentation using multiple generators or incorporate metamorphic relations or property-based testing.

\subsection{External Validity}
\textbf{Generalizability beyond benchmark Python functions.}
HumanEval, MBPP, and QuixBugs contain relatively small, self-contained functions with short specifications and unit-test harnesses. Real-world code review often involves larger codebases, hidden dependencies, stateful APIs, and non-functional requirements (performance, security, maintainability), therefore, the measured bias patterns and the effectiveness of the filter may not directly transfer to industrial settings.

% Our tasks are in Python and mostly algorithmic and utility-oriented, with natural-language specs that are relatively short and example driven. In practice, requirements can be ambiguous, incomplete, or distributed across documents (tickets, comments, design docs). Models’ over-correction may be different under longer, noisier, or domain-specific specifications, extending to other languages (e.g., Java/C++) and requirement styles is necessary to establish broader applicability.

\textbf{Model coverage and version drift.}
We evaluate five representative LLMs (three closed-source and two open-source). Closed-source models evolve rapidly, so measured rates may change as providers update models. For open-source models, alternative instruction-tuned variants or larger parameter scales may exhibit different bias profiles. While we report results for a diverse set of models, the findings should be interpreted as evidence about these specific model snapshots rather than immutable properties of the families.

\section{Conclusion}
In this paper, we revisits whether a model can reliably judge requirement conformance from a natural-language specification and an implementation without executing tests. Through a unified evaluation on three widely used benchmarks and five representative LLMs, we uncover a systematic reliability gap: models frequently reject correct implementations, and this over-correction tendency can become substantially worse as prompts are enriched with explanation and repair requirements. Our results demonstrate that prompt ``enhancements'' often act as a decision-boundary control rather than a guaranteed accuracy booster. In particular, explanation and fix-oriented prompting can sharply increase false rejections while reducing false acceptances, producing a pronounced trade-off that also translates into large operational costs in absolute FN/FP counts. Beyond verdict accuracy, we show that rationales are not consistently trustworthy: models can produce internally inconsistent explanations and exhibit high symptom-level but weaker cause-level fault awareness, implying that persuasive rationales may still be unfaithful or diagnostically misleading. To mitigate over-correction bias, we propose a Fix-guided Verification Filter that treats the model’s proposed fix as executable counterfactual evidence and validates the original vs.\ fix behavior in augmented tests with restrictions on parameters and specifications. This lightweight framework reduces false-negative rates across all tested models and benchmarks, with only modest increases in false-positive rates in some settings.

% highlighting the value of grounding LLM judgements in executable evidence rather than relying solely on more elaborate prompting. Overall, our findings caution against deploying LLMs as stand-alone requirement conformance judges in modern software development, we recommend treating prompt choice as cost-sensitive calibration, auditing explanation faithfulness separately from verdict correctness, and incorporating execution-guided verification whenever a model proposes repairs. 

\clearpage

\bibliography{sn-bibliography}% common bib file
%% if required, the content of .bbl file can be included here once bbl is generated
%%\input sn-article.bbl

\end{document}